\documentclass[%
 reprint,
 amsmath,amssymb,
 aps,
]{revtex4-2}
\usepackage{enumitem}
\usepackage{booktabs}
\usepackage{xcolor}
\usepackage{ragged2e} 
\usepackage{graphicx}
\usepackage{dcolumn}
\usepackage{bm}

\begin{document}

\preprint{Accepted for publication in Phys. Rev. D}
\title{Wavelet Scattering Transform for Gravitational Waves Analysis.\\ An Application to Glitch Characterization}

\author{Alessandro Licciardi}
\affiliation{%
 Department of Mathematical Sciences, Politecnico di Torino, 10129 Torino, Italy}\affiliation{
 INFN Sezione di Torino, 10125 Torino, Italy}%


\author{Davide Carbone}
\affiliation{
 Laboratoire de Physique de l’Ecole Normale Supérieure, ENS Université PSL, CNRS, Sorbonne Université, Université de Paris, Paris, France}%
\author{Lamberto Rondoni}
\affiliation{%
 Department of Mathematical Sciences, Politecnico di Torino, 10129 Torino, Italy}\affiliation{
 INFN Sezione di Torino, Torino, Italy}%

\author{Alessandro Nagar}
\affiliation{
 INFN Sezione di Torino, 10125 Torino, Italy}
 \affiliation{Institut des Hautes Etudes Scientifiques, 91440 Bures-sur-Yvette, France}


\begin{abstract}
Gravitational waves, first predicted by Albert Einstein within the framework of general relativity, were confirmed in 2015 by the LIGO/Virgo collaboration, marking a pivotal breakthrough in astrophysics. Despite this achievement, a key challenge remains in distinguishing true gravitational wave signals from noise artifacts, or "glitches," which can distort data and affect the quality of observations. Current state-of-the-art methods, such as the Q-transform, are widely used for signal processing, but face limitations when addressing certain types of signals. In this study, we investigate the Wavelet Scattering Transform (WST), a recent signal analysis method, as a complementary approach. Theoretical motivation for WST arises from its stability under signal deformations and its equivariance properties, which make it particularly suited for the complex nature of gravitational wave data. Our experiments on the LIGO O1a dataset show that WST simplifies classification tasks and enables the use of more efficient architectures compared to traditional methods. Furthermore, we explore the potential benefits of integrating WST with the Q-transform, demonstrating that ensemble methods exploiting both techniques can capture complementary features of the signal and improve overall performance. This work contributes to advancing machine learning applications in gravitational wave analysis, introducing refined preprocessing techniques that improve signal detection and classification.

\end{abstract}

\maketitle
\section{Introduction}
\label{sec:Introduction}
\justifying
Gravitational waves, predicted by Albert Einstein in the framework of general relativity \cite{einstein1915}, stand as one of the most significant discoveries of the past century. The experimental confirmation of these spacetime ripples, first achieved in 2015 by the LIGO/Virgo collaboration \cite{abbott2016observation}, has opened a new observational window into the universe, particularly in the context of multimessenger analysis of astrophysical data \cite{abbott2017multi}.
However, despite remarkable technical progress in gravitational wave detection, we are faced with a fundamental challenge: the identification of true signals in opposition to the so-called "glitches". Glitches represent interruptions or anomalies in the detected data that can distort signals of interest and compromise the accuracy of observations. These undesired phenomena may stem from a variety of sources, including environmental interference, instrumental errors, or data analysis issues. Establishing a taxonomy of such signals is fundamental to discerning between them and the true events; in fact, because of the non-gaussianity \cite{blackburn2008lsc}, standard preprocessing methods usually fail to remove glitches from the time series. \\
Most of the analysis of such spurious signals is based on spectrograms obtained through the Q-transform \cite{anderson1999time_qt_gw}, and even Bayesian methods make use of wavelet-like transforms \cite{cornish2015bayeswave}. More recently, time-frequency plots were collected in datasets, as Gravity Spy \cite{zevin2017gravityspy}, and used in the context of machine learning techniques used for image classification \cite{bahaadini2018machine_gravityspy}\cite{bahaadini2018machine_cnn__qtr}. Let us stress that the use of Q-transform is standard in the community; see, for instance, the Python libraries GWpy \cite{gwpy} and PyCBC \cite{nitz2020gwastro}. The result is that most of the statistical and machine learning analyses directly exploit images. \\
In other contexts, the analysis of complex signals like natural images or sounds has benefited from the use of more advanced preprocessing methods. Among all, the recently developed Wavelet Scattering Transform (WST) \cite{mallat2012group}\cite{bruna2013invariant} has sparked particular interest within the machine learning community. Contrary to standard wavelet or Fourier methods, the WST generates representations that enjoy stability properties with respect to small rototranslation and deformations. The basic aim of WST is to map elements which are close in the data space (e.g. a natural image or a sound) into representations close in the target, e.g. frequency, space. This request seems to be simple and fundamental for classification and clustering tasks, but standard methods used for temporal series, for instance Q-transform and short time Fourier transform, do not enjoy such stability properties. In this sense, WST has been shown to outperform standard preprocessing methods for 1D or 2D data, e.g. for music classification \cite{anden2014deep}, ECG tracks \cite{liu2020wavelet}, bioacustics \cite{10720021}, observational cosmology \cite{cheng2020new}\cite{valogiannis2022towards} and deep learning application \cite{oyallon2017scaling}.\\
In the present work we aim to evaluate the performance of WST as opposed to Q-transform in the context of Virgo data, and in particular for classification and clustering tasks on real glitches data. As expected, we show that the intracluster dispersion is critically reduced by WST; consequently, classification is made simpler and possible even with simple architecture, contrarily to the CNN needed for standard spectrograms obtained from Q-transform.       
In Section \ref{sec:WST} we review the theoretical definition of both preprocessing methods; in Section \ref{sec:Experiments} we present the experimental results which evidently shows a critical advantage in using WST in place of Q-transform. 
\section*{Related work}
The first papers on early Machine Learning techniques in the context of gravitational waves analysis trace back to the late 2010s. Initially, the focus was on unsupervised hierarchical learning \cite{mukherjee2006multidimensional}, early classification techniques \cite{mukherjee2010classification}, or simply the advantages of using GPUs for statistical analysis \cite{egri2008parallel}. A review \cite{hentschel2010machine} on Machine Learning for quantum measurement partially predicted the importance of ML applied to gravitational waves, and its broader application in Astrophysics was highlighted already in \cite{vanderplas2012introduction}\cite{way2012advances}.

With the advancements in machine learning techniques, the number of works applying ML to discriminate gravitational waves from noise artifacts increased as seen in   \cite{biswas2013application}\cite{powell2015classification}, until the milestone work on Gravity Spy dataset \cite{zevin2017gravityspy}. The existence of a benchmark dataset dramatically increased the scientific production in the area; see, for instance, the review \cite{cuoco2020enhancing}. The use of CNN follows state of the art in computer science \cite{gebhard2019convolutional}, and in spite of a usual tradition in the ML community, even open competitions regarding gravitational-data search were proposed, e.g. \cite{schafer2023first}.

Regarding preprocessing methods, the use of time-frequency spectrograms for gravitational waves traces back to \cite{anderson1999time_qt_gw}. Among all the proposed methods, the first work highlighting the use of wavelets \cite{mallat1999wavelet} and Q-Transform \cite{brown1991calculation} is \cite{chatterji2004multiresolution}; the two proposed methods become a sort of standard. For instance, Gravity Spy project made use of spectrograms and not of bare time-series. Just recently, the use of filtering methods based on ML has been proposed in \cite{yan2022generalized}, even as generative tool. {The coherent WaveBurst algorithm \cite{klimenko2008coherent}, widely used by the LVK collaboration, is a key wavelet-based approach for gravitational wave data analysis. More recently, the multi-resolution Wavescan algorithm \cite{klimenko2022wavescan} has been introduced to enhance the Q-transform by integrating multiple resolutions into a unified framework.}

Regarding Wavelet Scattering Transform \cite{mallat2012group}: its interpretability has been exploited in Physics Informed ML \cite{karniadakis2021physics} and for neural network analysis \cite{zhang2021survey}. The fields of application of such method are numerous; for the sake of the reader and with an eye to gravitational waves, we mention some regarding 1D signals: from music genre classification \cite{anden2014deep}, to EEG \cite{ahmad2017mallat} and ECG \cite{liu2020wavelet} in medicine, and in general to time-series analysis \cite{arouxet2021using}.

\section{Theory: Wavelet Scattering Transform}
\label{sec:WST}
In this section we present the standard definitions of Q-transform and WST; since the experimental analysis will focus on 1D data, for convenience of the reader we recall just the properties of WST in such field of application, even if it can be defined for data in higher dimension (as 2D or 3D images).
\subsection{State-of-the-art Representation: Q-Transform}
\begin{figure}[h]
    \centering
    \includegraphics[width=\linewidth]{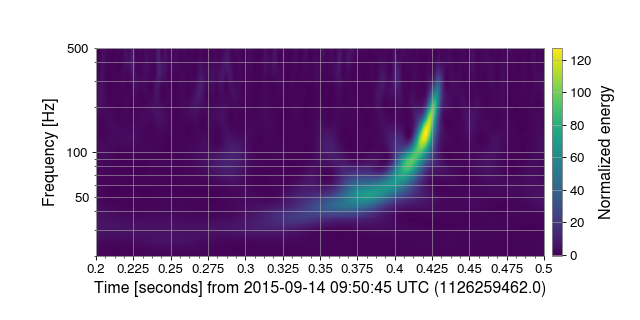}
    \caption{Spectrogram of GW150914 taken from GWpy repository \texttt{https://gwpy.github.io}, the first gravitational wave signal ever detected \cite{abbott2016observation}.}
    \label{fig:QT}
\end{figure}

The Q-transform, also known as the Constant Q-transform (CQT), is the most common technique for representing gravitational waves \cite{anderson1999time_qt_gw}\cite{chatterji2004multiresolution}. It is a time-frequency analysis technique that provides a representation of a signal in the joint time-frequency domain with a constant quality factor, denoted as $Q$.\\
Let $h(t)$ be a continuous-time signal, and consider a family of window functions $g_{Q}(t)$, such as Hann window functions, parameterized by a quality factor $Q$. The window functions have a constant bandwidth, centered at logarithmically spaced frequencies $\omega_k$.\\
Given a 1D signal $h(t)$, the Q-transform coefficient $H_k(t)$ at time $t$, for a specific frequency $f_k$ and factor $Q$ is obtained by the convolution:
\begin{equation}\label{q-transform_def}
H_k(t) = \int_{-\infty}^{\infty} h(\tau) \, g_{Q}(t - \tau) \, e^{- i f_k \tau} \, d\tau,
\end{equation}
To obtain the full Q-transform of the signal, we compute the Q-transform coefficients for a range of frequencies $f_k$ in the chosen frequency grid. The resulting Q-transform provides a time-frequency representation of the signal, where the time axis corresponds to the original signal's time domain, the frequency axis represents the logarithmically spaced $f_k$ and the color map refers to the module of the coefficients $H_k$. This kind of plot is referred to as spectrogram, see Figure \ref{fig:QT}.
The Q-transform is particularly useful for analyzing signals with varying frequency content, as it provides higher resolution at lower frequencies and lower resolution at higher frequencies. This logarithmic frequency scaling ensures that the Q-transform captures the details of signals with complex harmonic structures and varying frequency components \cite{brown1991calculation}. It, or its modification, enables the analysis of signals with complex frequency content and is widely applied in various domains, including audio processing, music analysis \cite{schorkhuber2010constant}\cite{schorkhuber2013audio}, and signal processing tasks \cite{sakar2019comparative}.\\
It is possible to delineate similarities and dissimilarities with the Short Time Fourier Transform (STFT) \cite{roberts1987digital_signal_processing}. On one hand, the STFT divides a signal into short, overlapping segments and applies the Fourier Transform to each segment. This results in a time-frequency representation where the time axis is preserved, but the frequency resolution is constant across all frequencies. The STFT provides good frequency localization at the expense of time resolution.\\
On the other hand, the Q-transform uses a family of window functions with constant bandwidths and logarithmically spaced frequencies, allowing for higher frequency resolution at lower frequencies and lower resolution at higher frequencies. The Q-transform provides a time-frequency representation with a constant quality factor, denoted as Q, which captures the details of signals with complex harmonic structures and varying frequency components. Therefore, the main difference between the STFT and the Q-transform lies in their frequency resolution characteristics. The STFT offers a constant frequency resolution but sacrifices time resolution, while the Q-transform provides variable frequency resolution with a constant quality factor, allowing for a better representation of signals with varying frequency content, such as gravitational waves \cite{anderson1999time_qt_gw}.

\subsection{Wavelet Scattering Transform for time series}
\begin{figure}
    \centering
    \includegraphics[width=\linewidth]{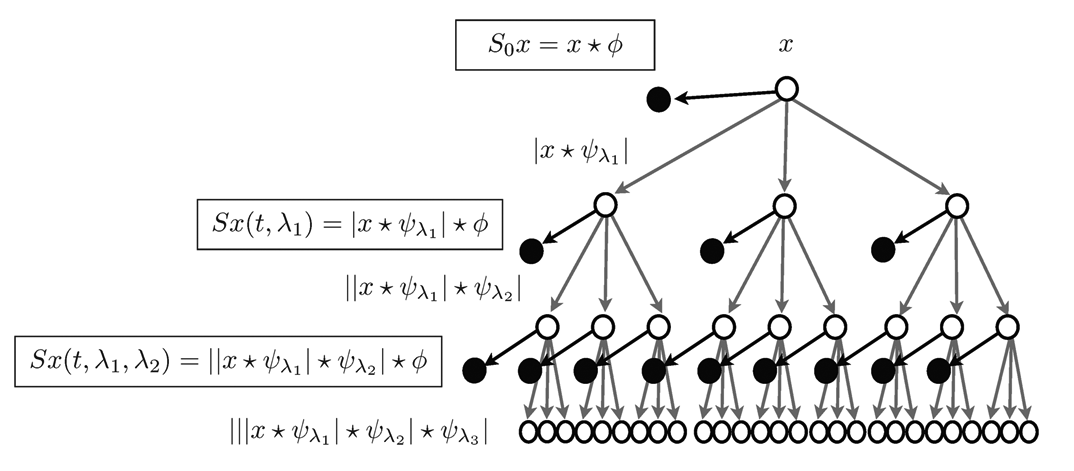}
    \caption{Visualization of Wavelet Scattering Transform as an iterative process, from \cite{anden2014deep}. In their notation the signal is $x(t)=h(t)$ the path $p$ at depth $m$ is explicited in parentheses as a tuple $(\lambda_1,\dots,\lambda_m)$. }
    \label{fig:tree-wst}
\end{figure}

The Wavelet Scattering Transform (WST) \cite{mallat2012group} is a mathematical operator that allows to obtain for a given signal a stable and invariant representation. In particular, under suitable hypothesis \cite{bruna2013invariant}, the representation is translation invariant, stable to additive noise, i.e. it is non-expansive, and stable to deformations; this property is mathematically expressed in its original derivation as Lipschitz-continuity under the action of $C^2-$diffeomorphisms.  Utilizing a representation operator with these properties in a machine learning scenario could significantly reduce the computational effort required in training classification algorithms \cite{brunaphd}.
\\
The construction of the WST on one dimensional signals relies on properly defining a family of wavelet functions, starting from a mother wavelet $\psi \in L^2(\mathbb{R},dx)$. Let $a>1$ be a scalar. A family of wavelet can be defined as 
\begin{equation}\label{waveletfunctions}
    \psi_{\lambda}(t)=\lambda^{-1}\psi(\lambda^{-1}t)\;\;\; \lambda\in a^\mathbb{Z}\,,
\end{equation}
where $\lambda=a^j$. In practice, for greater values of $j$ the wavelet has a larger support in time domain, hence a smaller bandwidth in frequency domain. In the usual definition of WST, they define $A\in\mathbb{N}$ such that $a=2^{1/A}$; this will play a role of a hyperparameter.  Furthermore, we fix a maximum {\it depth} $J \in \mathbb{Z}$, corresponding to a set of allowed scaling operators $\lambda$, i.e. belonging to the set
\begin{equation}
\label{depth}
    \Lambda_J=\{\lambda \in a^\mathbb{Z}\,: |\lambda|=a^j\leq 2^J\}\,.
\end{equation}
The couple $(J,A)$ contains the hyperparameters of the WST, analogously to $Q$ for the Q-transform.\\
In order to establish the desired stability properties, the WST exploits the non-linearity introduced by the so-called \textit{propagator operator} \cite{mallat2012group}. The propagator operator over a path $(\lambda_1,\dots,\lambda_m)$, with $\lambda_i\in\Lambda_J$, applied to a given signal $h(t)$, is defined as
\begin{equation}\label{propagator}
    U[p]h(t)=U[\lambda_m]\dots U[\lambda_1]h(t)\,,
\end{equation}
where, denoting with $\star$ the convolution, $U[\lambda]h(t)=|\psi_\lambda \star h(t)|$.
Therefore the propagator operator cascades convolutions and moduli -- this attempt tries to emulate the structure of a layer of a convolutional neural network \cite{mallat2016understandingCNN}.
The WST of a signal $h(t)\in L^1(\mathbb{R},dx)$ over a path $p=(\lambda_1,\dots,\lambda_m)$, with $\lambda_i\in \Lambda_J$, is defined as
\begin{equation}\label{WST_def}
    S_J[p]h(t)=U[p]h(t)\star\phi_J(t)\,,
\end{equation}
where $\phi_J(t)$ is a low-pass filter, scaled according to the depth parameter $J\in\mathbb{Z}$. Namely
\begin{equation}
S_J[p]h(t)=|\psi_{\lambda_m}\star|\dots|\psi_{\lambda_2}\star|\psi_{\lambda_1}\star h||\dots|\star\phi_J(t)\,.
\end{equation}
For the conducted experiments we couple \textit{Morlet wavelets} with a Gaussian low-pass filter \cite{mallat1999wavelet}.\\
As in \cite{mallat2012group}, introducing the path set up to length $m$, $\Lambda_J^m=\{(\lambda_1,\dots,\lambda_m):\,|\lambda_i|=a^j\leq 2^J\}$, it is possible to define the induced norm of the scattering operator over the set $\mathcal{P}_J=\bigcup \Lambda_J^m$, i.e.
\begin{equation}\label{scattering_norm}
    \|S_J[\mathcal{P}_J]h\|=\sum_{p\in \mathcal{P}_J}\|S_J[p]h\|
\end{equation}
where $\|\cdot\|$ stands for the $L^2-$norm. For fixed $J$ and $A$, and given the definition of $\Lambda_J^m$, the WST can be graphically visualized as an iterative tree process, see Figure \ref{fig:tree-wst}.
One could be concerned about the depth requested in practice, but in different experiment \cite{brunaphd} it has been showed that just $2$ or $3$ layers of WST are sufficient to represent around $98\%$ of the energy of the signal. Indeed the energy of each layer, i.e. $\|U[\Lambda_J^m]\|$, is empirically observed to rapidly converge to zero.
The norm induced by the WST is \textit{non-expansive}, therefore the representation operator is stable to additive perturbation. Hence, for any signal $h\in L^2(\mathbb{R},dt)$ and any perturbed version $h'=h+\epsilon$, it holds
\begin{equation}\label{non_exp}
    \|S_J[\mathcal{P}_J]h'-S_J[\mathcal{P}_J]h\|\leq \|h'-h\|=\|\epsilon\|\,.
\end{equation}
Moreover through the WST operator is possible to achieve translation invariance \cite{bruna2013invariant}, i.e. for any signal $h(t)\in L^2(\mathbb{R},dx)$ and any translated version $h'(t+a)$ it holds
\begin{equation}\label{trans_inv}
    \lim_{J\to\infty} \|S_J[\mathcal{P}_J]h'-S_J[\mathcal{P}_J]h\|=0\,.
\end{equation}
Further theoretical results \cite{mallat2012group} show that, when hypothesis are met, the representation is stable to continuous domain stretches and desplacement fields.
These properties are expected to become particularly relevant when studying gravitational waves data, where the information is highly affected by noise, due to the high sensitivity of the interferometers.\\ We stress that centering the peak of the signal before preprocessing is a core problem is glitch analysis; since WST is less sensible to time shifts and deformations, we expect it to be helpful against this issue.
\begin{figure}[h]
    \centering   \includegraphics[width=0.49\linewidth]{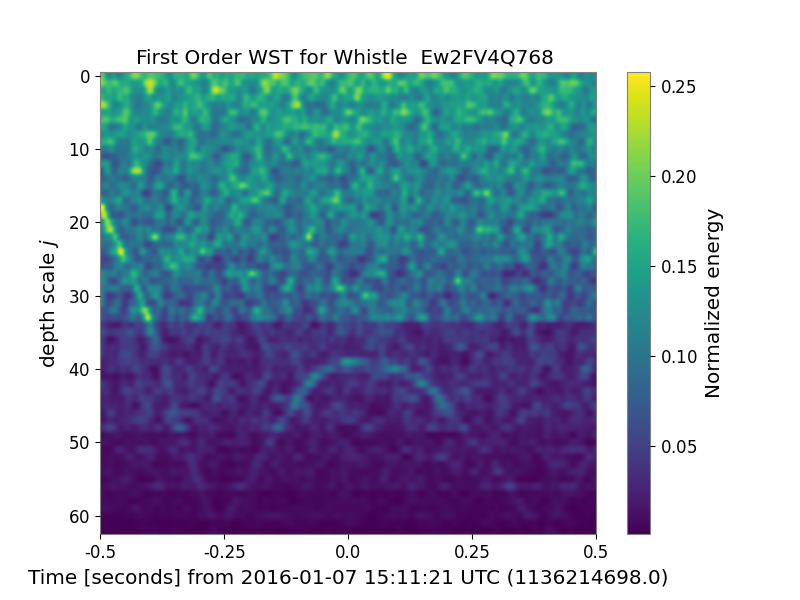}\includegraphics[width=0.49\linewidth]{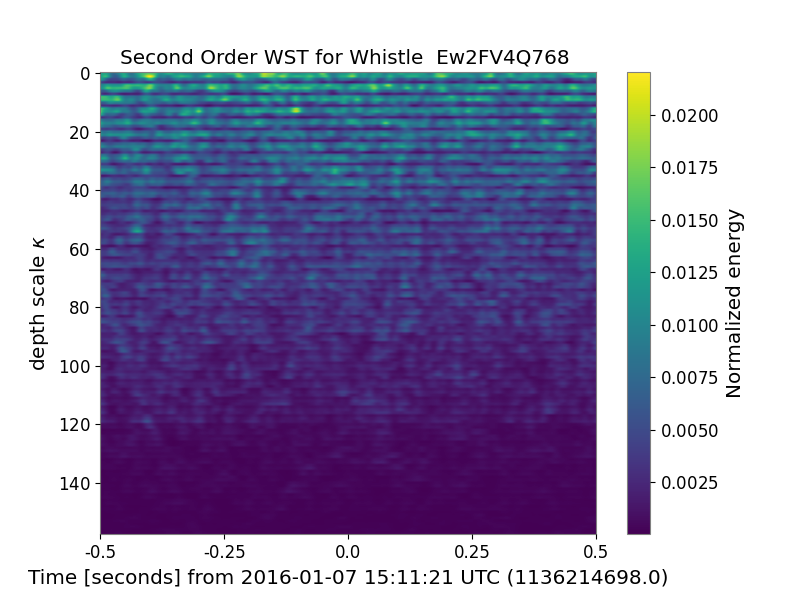}
    \caption{First and second order WST for the glitch with ID \texttt{Ew2FV4Q768}. The plots are not normalized and no customized color map is applied.}
    \label{fig:wst-notnorm}
\end{figure}
\begin{figure*}[t]
    \centering
    \includegraphics[width=0.32\linewidth]{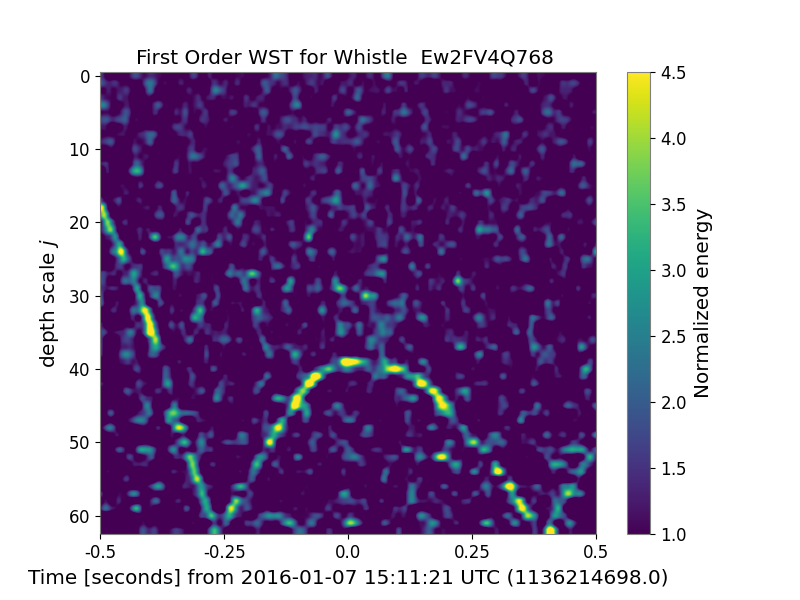}
    \includegraphics[width=0.32\linewidth]{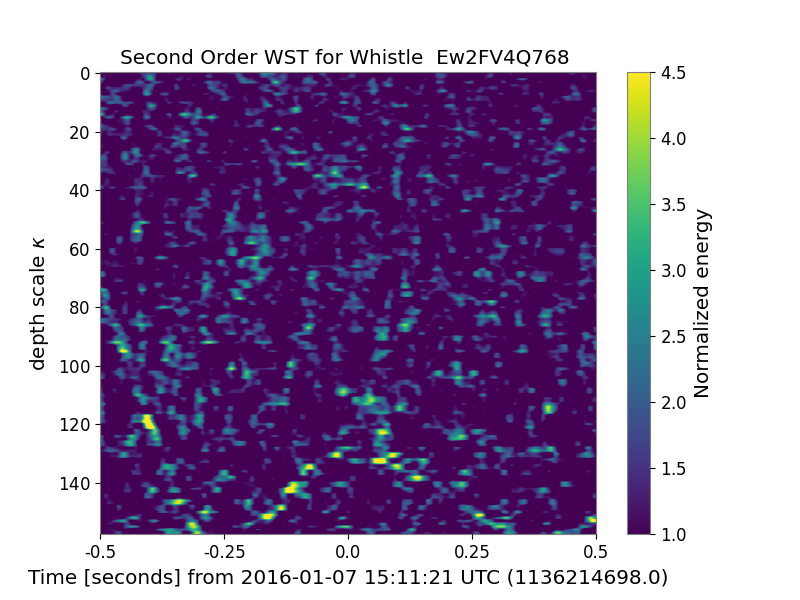}
    \includegraphics[width=0.32\linewidth]{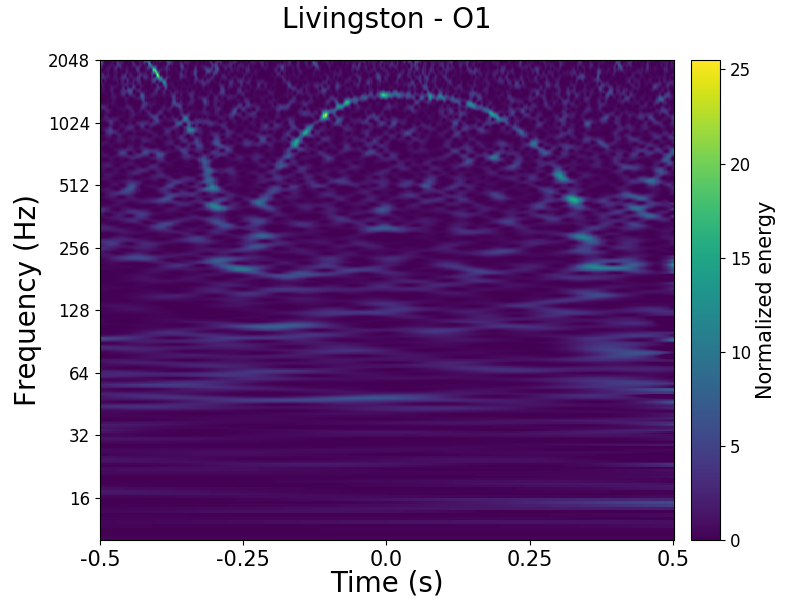}
    \caption{{\it Left and Center:} First and second order WST for the glitch with ID \texttt{Ew2FV4Q768}. The plots are normalized row-wise and the original signal is not downsampled to 2048 Hz. For the sake of visualization, we used $J,A=(6,16)$ and the customized color map \eqref{cmap} is applied. }
    \label{fig:comp_wst_q}
\end{figure*}
\subsection{{Comparison Between Q-Transform and WST}}
{The main takeaway is that in several scenarios, such as 1D time series classification, machine learning models can achieve state-of-the-art performance with smaller architectures when using WST \cite{bruna2013invariant}. These models are consequently easier to train due to a reduced number of trainable parameters. Mathematically, this advantage arises from WST’s ability to map data into a feature space that enhances separability by satisfying the following properties:
\begin{enumerate}[label = \roman*]
    \item stability to additive noise
    \item local translation invariance
    \item stability to small continuous time warping
\end{enumerate}
By construction, WST satisfies these properties, theoretically ensuring better data separation. In contrast, the Q-transform and standard wavelet transforms do not generally satisfy these properties, which can result in less robust feature representations—for example, small additive noise may significantly alter the feature space representation.
The key structural differences arise from WST’s non-linearity and recursive nature. While the Q-transform (Eq. \eqref{q-transform_def}) is a linear operator based on convolution with a specific window function, linearity alone prevents it from achieving properties (i), (ii), and (iii). In contrast, first-order WST introduces non-linearity through the modulus operation and applies an averaging step using a low-pass filter, leading to improved feature stability and separability. As far as the analysis led in our study is concerned, other differences are in the hyper-parameters, such as the choice of the wavelets, or the frequency resolution scale.
}
\section{Training and test datasets set-up}
In this section, our objective is to conduct a comprehensive comparison of the data analysis between the Q-transform and the WST. We reconstruct and explicitly describe the pipeline for the preprocessing adopted in \cite{bahaadini2018machine_gravityspy}. Notably, the application of WST as a theoretical tool is already prevalent in diverse fields such as cosmology \cite{valogiannis2022towards} and field theory \cite{marchand2022wavelet}. As a widely acknowledged principle in the literature \cite{bruna2013invariant}, WST is preferable to STFT methods when their performances are comparable, primarily due to the invariance properties that facilitate cross-signal interpretation.  

\subsection{Gravity Spy 1.1.0}
\begin{figure}[h]
    \centering
    \includegraphics[width = 0.8\linewidth]{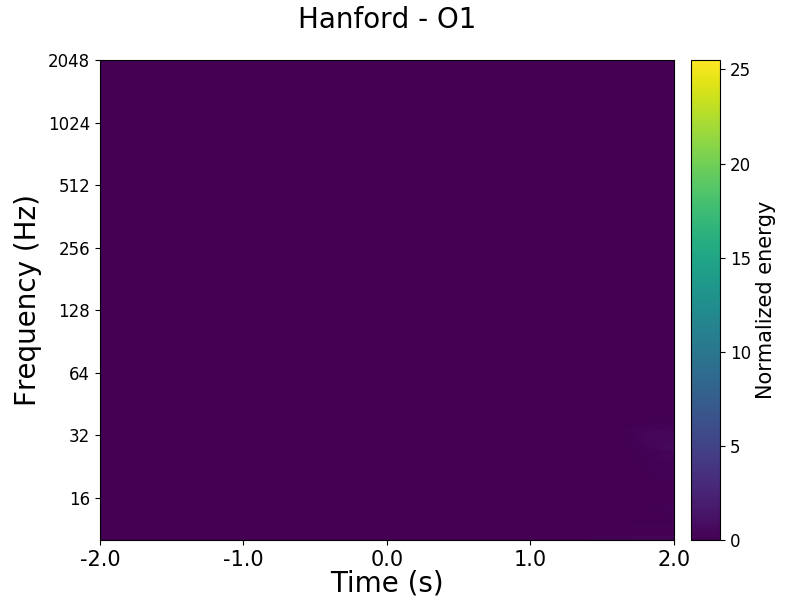}
    \caption{Image corresponding to the file \texttt{H1\_rymrZrzFCx\_spectrogram\_4.0.png} in the dataset of Gravity Spy 1.1.0.}
    \label{fig:nan}
\end{figure}

\begin{figure}[h]
    \centering
    \includegraphics[width = .8\linewidth]{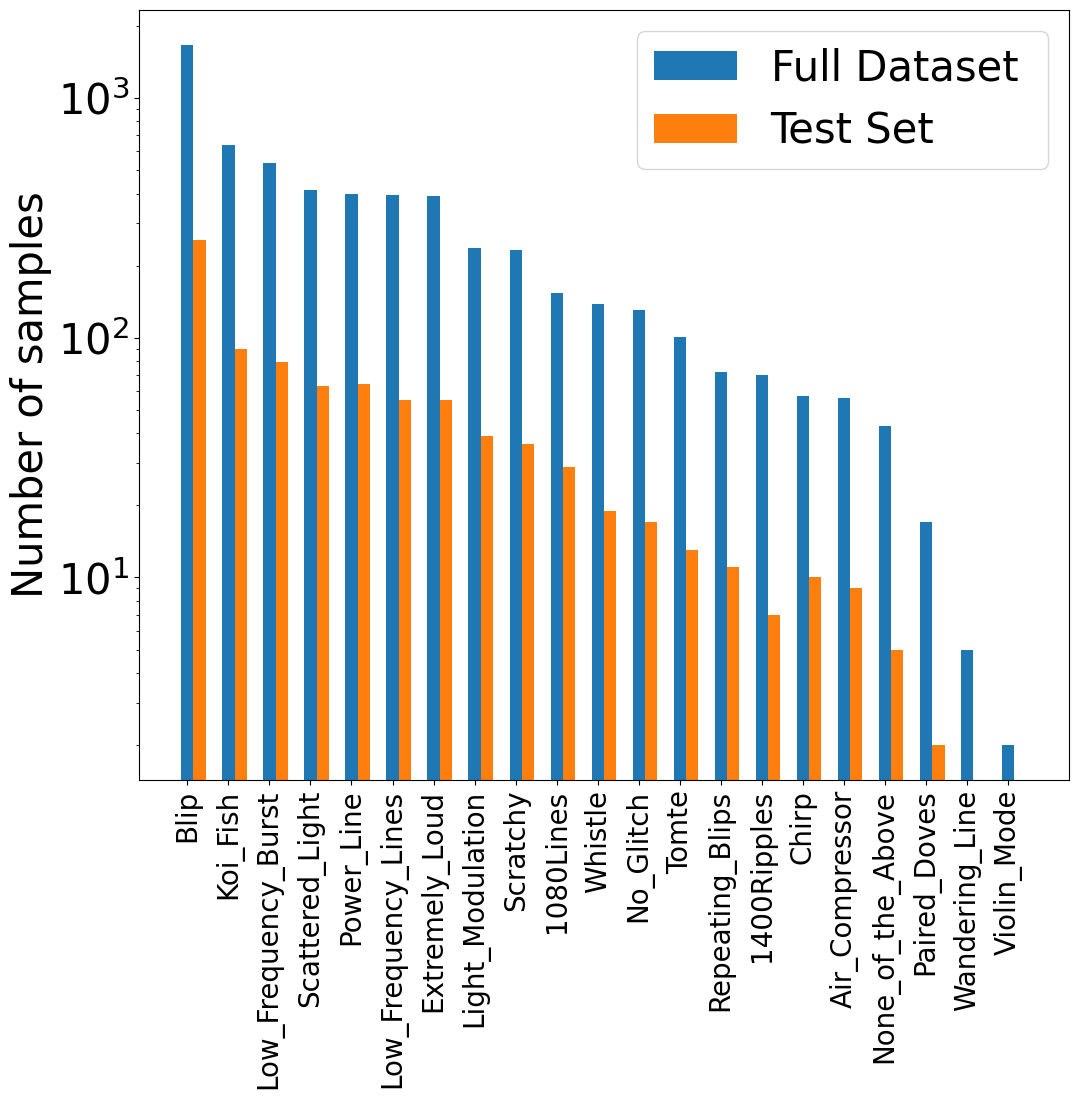}
    \caption{Number of samples per class after eliminating the time series containing NaN values and the four observations sampled at $4096$ Hz, in log-scale and sorted in decreasing order. The dataset is very imbalanced: \textit{Blip} is the most represented glitch and contains $1668$ instances, while the smallest, \textit{Violin Mode}, has just $2$ samples in the training set.}
    \label{fig:distri}
\end{figure}

In this study, we use Gravity Spy 1.1.0 \cite{bahaadini2018machine_gravityspy} as a foundational dataset for our research. The database is public and available at \url{https://zenodo.org/records/1486046} and contains a collection of glitches, alongside with a detailed description of its content. For readers' convenience, we report the main features of the available material. From the highlighted source, it is possible to download a collection of images containing the spectrograms computed with Q-transform. Each image is  identified by an ID and it is possible to associate the corresponding time series using the available csv file. Then, every time series can be downloaded using the method \texttt{fetch\_open\_data()} from GwPy library. \\
Our first objective was to recompute the presented spectrograms as sanity check using an independent pipeline. Before proceeding to the details of the preprocessing, some time series present in the csv file contains NaN entries when downloaded because gated before the public release, and so the corresponding spectrograms in the public dataset are empty; as for instance in the glitch with id \texttt{rymrZrzFCx} in the class \textit{Extremely Loud} (see Figure \ref{fig:nan}). Some signals appear to have a sample rate of 4096 Hz; removing such time series is necessary since the \texttt{q\_transform} method in GwPy requires higher sample rate to build a spectrogram with maximum frequency 2048 Hz (see GwPy documentation for detailed information).  \\The first passage is then to detect and remove such time series from the dataset. In the present work, we consider the dataset as presented in the original paper, composed by 22 classes; however, after the removal of NaN instances, the number of classes reduces to 21 (see Figure \ref{fig:distri}). 

\begin{figure*}[t]
    \centering\includegraphics[width=0.33\linewidth]{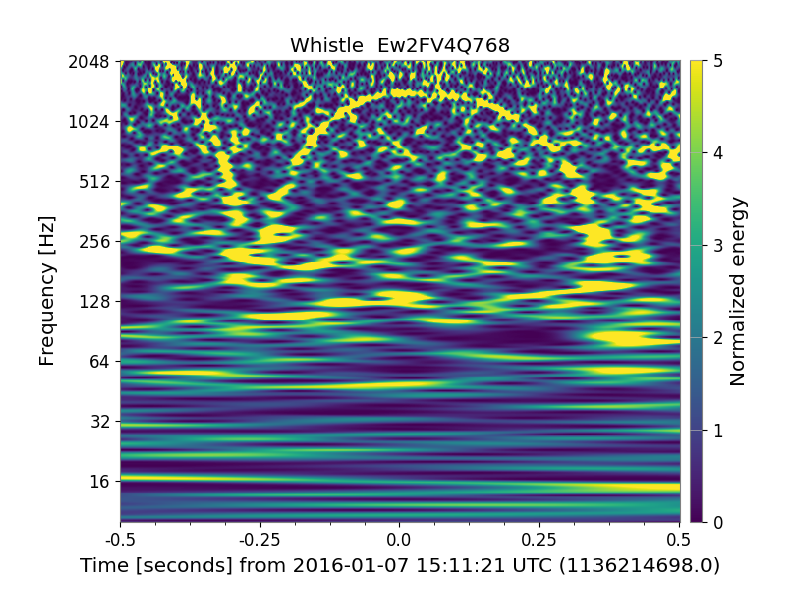}\includegraphics[width=0.33\linewidth]{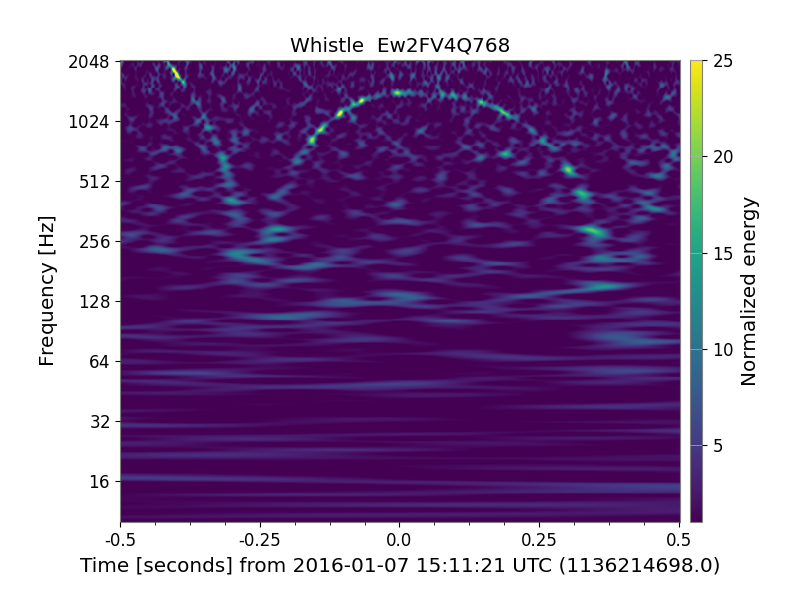}
    \includegraphics[width=0.33\linewidth]{images/L1_Ew2FV4Q768_spectrogram_1.0.png}
    \caption{Spectrograms corresponding to the same signal (id: \texttt{Ew2FV4Q768}) with different values of $c_{min}$ and $c_{max}$ {-- the same color map and corresponding image in Gravity Spy are used for consistency }. {\it Left:} $c_{min}=0$ and $c_{max}=5$. {\it Center:} $c_{min}=1$ and $c_{max}=25$. {\it Right:} image taken from Gravity Spy dataset.}
    \label{fig:cmapdiff}
\end{figure*}

\subsection{Data processing}
We download the time series indexed in the file \texttt{trainingset\_v1d1\_metadata.csv}, with a window of $\pm 2\, s$ centered in the \texttt{event\_time} attribute of each signal. As briefly discussed in the previous subsection, we then seek for NaN entries in the time series; hence, we remove such instances from the list of time series. The second step is whitening, performed with the corresponding \texttt{whiten()} method in GwPy library. After these introductory passages, we have raw time series with sample rate of $16384$ Hz. We now separately present the pipeline for Q-transform and spectrogram generation, as inferred from \cite{bahaadini2018machine_gravityspy}. Subsequently, we present the procedure for WST.  \\
\paragraph{Q-transform.} We use the method \texttt{q\_transform()} present in GwPy library to compute the spectrogram from the \texttt{TimeSeries} objects we obtained from the previous steps. As for the arguments of such methods, we followed \cite{bahaadini2018machine_gravityspy} which adopts: $\textit{qrange}=(4,64)$ for the range of $Q$ search. We recall that the obtained spectrogram is in fact the result of a tiling procedure, as described in GwPy documentation. Then,  $\textit{fres}=470$ and $\textit{tres} = 2*\delta_t / 570$ in order to obtain a spectrogram of dimension $470\times 570$, where $\delta_t=2\, s$ is the semi-length of the time window. Coherently, the \textit{outseg} is chosen to be the segment centered in the \texttt{event\_time} with with size $2\delta_t$. Subsequently, we fix $\textit{frange}=(10, 2048)$ for the frequency range, $\textit{whiten}=False$ since already performed and $\textit{logf}=True$ to span frequency range in logarithmic scale. At last, we normalized the obtained spectrogram with respect to the mean using $\textit{norm}=\text{'mean'}$. We stress again that by construction of the GwPy Q-transform, we are forced to preprocess signal with high sample rate even if the frequency axis is capped at 2048 Hz. \\
At this point, as suggested in the original paper, the image is reshaped to the size $140\times170$; we use the method \texttt{Resize} from Pytorch library for this task. The final step is the choice of the normalization, that is of the color map present in the effective plots. We stress that every step modifying the image is fundamental since the final task will be image classification; there is no information, when performing a machine learning task, about the original time series. For instance, the classification accuracy could be dramatically different for a classifier trained on image having linear, and not logarithmic, frequency scale. The same occur for the choice of the color map, since the spectrogram can appear really different.
Coherently with the attribute \texttt{cmap} present in \texttt{Matplotlib} package, we fix an upper and lower cap such that $0<c_{min}<c_{max}$ and we remap the value of normalized energy $p$ of each pixel using the following piecewise function:
\begin{equation}
\label{cmap}
f(p)=
    \begin{cases}
        c_{max}\quad \text{if}\quad p\geq c_{max}\\
        p \quad \text{if}\quad c_{min}<p< c_{max}\\
        c_{min}\quad \text{if}\quad p\leq c_{min}\\
    \end{cases}
\end{equation}
This function represent a capped linear color map. In Figure \ref{fig:cmapdiff} we show how different values of the bounds can dramatically impact the appearance of the image. Moreover, we also present the comparison with the corresponding one present in the image folder of the downloaded dataset. As a side note, every image is plotted with a forced squared aspect ratio; such choice is maintained throughout the whole paper.\\

\paragraph{Wavelet Scattering Transform.} For each signal, we compute the Wavelet Scattering Transform (WST) up to the second order on a time window of $4\, s$ centered in the event time. Since the Q-transform cap the frequency at 2048 Hz, we decided to downsample the signal at that frequency before computing the WST. We choose $(J,A)=(7,7)$ as hyperparameters for both orders.  The zeroth-order WST, which provides no informative content, was excluded from the analysis. As far as the dimensions of the resultant images is concerned, for the chosen configuration the dimensions for the first and second order are 42$ \times$128 and 114$ \times$128, respectively.\\

We plot in Figure \ref{fig:wst-notnorm} the unnormalized first and second order WST for a chosen signal before downsampling. Let us focus for a moment on the meaning of such plots: while the content of the time axis is evident, we recall that each depth parameter, for the first order, correspond to a precise bandwidth of the wavelet used for the convolution operation. For the second order, we use an identification rule $(j_1, j_2)\to \kappa$ to map the couples of admitted depths in a single value. Their definition was introduced in Section \ref{sec:WST}.\\
As evident from Figure \ref{fig:wst-notnorm}, a normalization appears to be necessary, even at first glance. Analogously with Q-transform, we normalize with respect to the mean; the main dissimilarity is that such normalization is performed row-wise, that is for every value of $j$ and $\kappa$. Despite the fact that the use of the color map seems to be suggested also for WST (e.g. Figure \ref{fig:comp_wst_q}), we noticed that better classification performance (see next Section) are obtained without the use of color map. We stress as this is not a limitation; on the contrary, classification results based on the choice of the color map, as for Q-transform, can be problematic since based on the human eye. 

In conclusion, for the sake of a direct comparison with Q-transform, we reshape the size $140\times170$. We stress as for WST we are actually performing an interpolation, since the original dimension along depth axis is lower than 140.
\subsection{Training and Test Datasets}
To ensure a rigorous validation experiment, the cleaned and preprocessed dataset was split into two distinct subsets. Specifically, following the original split proposed in the csv file, 70\% of the data samples were allocated for model's training, while the remaining 30\% were equally split in validation (15\%) and test (15\%). We adopted their same split for the sake of a precise comparison. As per standard practice, generalization capability is assessed using the test set, which is strictly excluded from the training process of the classifiers.
\subsection{Software and Computational Resources}
The Q-transform is computed utilizing the GwPy library, whereas the Wavelet Scattering Transform (WST) is performed using the \texttt{Kymatio} library \cite{andreux2020kymatio}. The training of the classifier is executed on the high-performance computing (HPC) facilities of Department of Mathematical Sciences (DISMA) at Politecnico di Torino.
\subsection{Classifier}
For the sake of the present work, we use a simple Random Forest Classifier, as taken from \texttt{CuML} Python library. We use it separately on Q-transform, first order and second order of WST. Then, coherently with the proposal of \cite{bahaadini2018machine_gravityspy}, we apply a merge method to use the best of each preprocessing. In particular, we readopt the {\it hard merge}, which corresponds to build a convex combination of the predicted probability vector on the validation set, i.e. for $p_1$ and $p_2$ vectors of probability predictions
$$
p_{merge}=\alpha p_1 + (1-\alpha) p_2
$$

The test set is {\it never used} for hyperparameter selection; then we perform a grid search on the hyperparameter $0\leq\alpha\leq1$ in order to maximize the accuracy. We perform such merge procedure firstly between first and second order of WST, that we name for the sake of brevity "S1+S2"; then we define a second convex combination between the probability predictions using "S1+S2" and using the Q-transform. 
An immediate follow up for our analysis is the use of CNNs, as proposed in \cite{bahaadini2018machine_cnn__qtr}.

\section{Experimental Results}
\label{sec:Experiments}

We present in Table \ref{tab:results_table} the accuracy result on the test set for the classifiers on the first order WST, on the second order WST, on Q-transform, on "S1+S2" and on the total merge of WST + Q-transform. 

Evidently, for the dataset in study there is no statistical advantage in the use of Q-transform over WST. However, the possibility of having an independent preprocessing methods allows to combine them with ensemble methods, such for example an hard merge. The final result we obtain is an improvement of {3\%} of the classification accuracy on the test set.\\
In details, second order WST {slightly increases the performance of the first order of the WST} but a detailed grid search on $(J,A)$ could reveal more insights about this fact. Combining the two orders, i.e. in "S1+S2", leads to {an notable improvement with respect to the state-of-the-art. Moreover, as depicted in Table \ref{tab:ablation}, where we compare Q-transform and WST by changing the color-scale of the representation, it is possible to see that the WST consistently performs better than the Q-transform.} 
However, when the "S1+S2" classifier is merged with the Random Forest trained on Q-transform we manage to obtain an improvement of $2\%$ in accuracy.
\begin{table}[h]
  \centering
  \begin{tabular}{>{\centering\arraybackslash}m{5cm} >{\centering\arraybackslash}m{3cm}}
    \toprule
    \textbf{Preprocessing Method} & \textbf{Accuracy (\%)} \\ \midrule
    First Order WST               & {88.11}                  \\ 
    Second Order WST              & {68.65}                 \\ 
    {Total} WST      & {88.57}             \\ 
    Q-transform                   & {87.41}                  \\ 
    {WST + Q-transform}                   & {\textbf{91.03}}         \\ \bottomrule
  \end{tabular}
  \vskip 1.5em
  \caption{Test set accuracy for different preprocessing methods. {In this case we did not applied a colormap on the WST, while the Q-transform had been rescaled to $(0,25)$, as in \cite{bahaadini2018machine}. With these hyper-parameters the WST improves the performance of around 1\%. It is interesting to see that the best performance had been achieved by combining WST and Q-transform, improving the test accuracy to 91.03\%.} }
  \label{tab:results_table}
\end{table}

\subsection{Ablation on the Colour Scale}
In this section we show that WST method is robust with respect to changes in the color scale, while, on the other hand the Q-transform method is more sensitive to this further pre-processing step. In particular, in accordance to what reported in \cite{bahaadini2018machine} Q-transform achieve the best results when its color map is changed to $(0,25)$, while WST has its best performance without imposing a color-scale. In Table \ref{tab:ablation} we present the results obtained with different color scales.
\begin{table*}[t]

\renewcommand{\arraystretch}{1.5}
\centering
\begin{tabular}{c|c|c|c|c|c}

    \toprule 
       ${c_{max}}$& \textbf{First Order WST} & \textbf{Second Order WST} & \textbf{\textcolor{black}{Total} WST} & \textbf{Q-transform} & \textbf{\textcolor{black}{WST + Q-transform} } \\
       \midrule
1  &   81.35 \%            &     63.75 \%             &  81.47\%                    &  75.17\%            &       82.28\%       \\
25 &     88.11\%              &   68.65\%        &     88.57\%  &      87.41\%   & \textbf{91.03\%}      \\

\textbf{None}   &          88.11\%        &         68.65\%          &                         88.57\%  & 87.30\%    & 90.56\% \\
\bottomrule
\end{tabular}

\caption{{The table presents the test accuracy of various pre-processing methods as correlated with differing maximum values of the color scale. The WST color scale has been adjusted using a colormap within the range $(-c_{max}, c_{max})$, as the representation includes negative features that are crucial to the glitch detection process. Conversely, the Q-transform has been adjusted to a range of $(0,c_{max})$. Notably, the WST outperforms the Q-transform, with an average improvement of 1\%. However, when implementing the combined approach of WST and Q-transform, referred to as WST + Q-transform , there is an enhancement of 2 to 3\% compared to the Q-transform alone.}}\label{tab:ablation}
\end{table*}
\subsection*{Computational Cost}
Aside from the possible use in ensemble methods, a fundamental comparison between Q-transform and WST regards the computational cost and speed. Actually, this provides a further motivation for the adoption of WST when possible. In fact, the possibility to perform real time analysis of the signals is a fundamental problem in gravitational wave detection and processing \cite{dax2021real}\cite{gunny2022hardware}. 
The comparison between GwPy and Kymatio on CPU seems to show that the computational cost is similar. However, WST performed on GPU clearly outperforms Q-transform in term of computational time. We stress as the GPU we used are not state-of-the-art in term of performance, and so that such result can be possibily improved.\\  Moreover, there is another clear advantage in using WST on GPU: the latter can be computed by batch, while Q-transform has to be computed sample by sample. The benefit in terms of computational time is critical. {Specifically, the computation times are \(0.211 \pm 0.003\) seconds per sample for the Q-transform (GwPy, CPU), \(0.197 \pm 0.007\) seconds for WST (Kymatio with NumPy, CPU), and \(0.063 \pm 0.002\) seconds for WST (Kymatio with PyTorch, GPU), indicating that the GPU-based WST implementation is approximately \(3.13\times\) faster than the CPU-based WST and \(3.35\times\) faster than the Q-transform. The total number of considered samples is $100$ and the standard error is used to quantify the uncertainty.} For instance, on GPU for a batch of $100$ samples the computational time is $0.063$, which is basically equivalent to the time needed per one single sample. Such results represent a motivation towards the use of WST is glitch classification, or in general in time series analysis, especially when a machine learning task, and so batching, is involved. \\
{Apart from the computational time associated with the two pre-processing methods, the cost of the two methodologies is comparable. In particular, the implementation of the WST depends on the fast scattering algorithm \cite{anden2014deep}, which utilizes a highly efficient embedding for convolutions. The computational cost of the first-order WST is $\mathcal{O}(JAT\log T)$, where $T$ represents the number of time stamps of the signals, and $J,A$ are the WST hyper-parameters. Conversely, extracting features from the second order of the WST incurs a cost of $\mathcal{O}(J^2 A T \log T)$ due to an increase in the number of scattering coefficients. While documentation on the Q-transform implementation of GwPy is limited, it is plausible that its cost can be represented as $\mathcal{O}(QM T \log T)$, where $Q$ denotes the number of tiles comprising the Q-tile, and $M$ signifies the number of windows. We assessed the efficiency of the two representations by comparing the number of elementary operations and functions required to transform the same signal, utilizing the \texttt{cProfile}, which resulted in 258,761 for the Q-transform versus 14,374 for the WST. This comparison highlights the efficiency of the WST.}
\section{Conclusions}
\label{sec:Conclusions}
In this study, we conduct a thorough comparison between the Q-transform, the current leading preprocessing method for gravitational wave analysis, and the Wavelet Scattering Transform (WST), a recently developed mathematical tool for data preprocessing. Our work reveals how, for the dataset available, the two preprocessing methods leads to analogous results in term of classification accuracy. Notably, WST seems to bypass the issue of the color scaling of the spectrogram, which is potentially subtle since based on human eye. Plus, we stress as the fundamental advantage of WST stems in a notable reduction in computational time, surpassing Q-transform by more than one order of magnitude. Moreover, WST is completely compatible with \texttt{TensorFlow} and \texttt{Pytorch}, making it naturally applicable in deep learning applications. We believe that a deeper investigation of the hyperparameters of WST, coherently with a more systematic contruction of the dataset, could possibly provide even better motivation for the use of WST. Moreover, being WST a different independent preprocessing method, we show how one can improve the classification accuracy with a merge method, considering the best of two worlds. \\
Looking ahead, future directions necessitate testing WST on other collections of data of glitches and gravitational waves to further validate our results. While we anticipate no conceptual challenges, a larger dataset would enable more comprehensive testing in classification, potentially achieving state-of-the-art performance with simplified architectures as SVM or even CNN. 
\vspace{6pt} 


\section*{Acknowledgements}
We are grateful to L.~Asprea, F.~Legger, F.~Sarandrea and S.~Vallero
for discussions and help in getting access to Virgo data at the 
beginning of these project.
A.L., D.C. and L.R. worked under the auspices of Italian National Group of Mathematical
Physics (GNFM) of INdAM. A.L. is part of the project PNRR-NGEU which has received funding from the MUR – DM 117/2023. 
\vfill

\bibliography{apssamp}

\end{document}